\documentclass[pre,twocolumn,superscriptaddress,showpacs]{revtex4}

\usepackage{amsmath}
\usepackage{amsfonts}
\usepackage{amssymb}
\usepackage[pdftex]{graphicx}
\usepackage{verbatim}
\usepackage{subfigure}
\usepackage{sidecap}
\usepackage{color}
\usepackage{url}

\newcommand{\ER}{Erd\H{o}s-R\'{e}nyi }

\begin{document}

\title{Multiplexity-facilitated cascades in networks}

\author{Charles D.\ Brummitt}
\affiliation{Department of Mathematics and Complexity Sciences Center, University of California, Davis, CA 95616}
\author{Kyu-Min Lee}
\author{K.-I.\ Goh}
\email{kgoh@korea.ac.kr}
\affiliation{Department of Physics and Institute of Basic Science, Korea University, Seoul 136-713, Korea}

\date{\today}

\begin{abstract}
Elements of networks interact in many ways, so modeling them with graphs requires multiple types of edges (or network layers). Here we show that such multiplex networks are generically more vulnerable to global cascades than simplex networks. 
We generalize the threshold cascade model [D.~J.~Watts, Proc.~Natl.~Acad.~Sci.~U.S.A.~{\bf 99}, 5766 (2002)] to multiplex networks, in which a node activates if a sufficiently large fraction of neighbors in any layer are active. 
We show that both combining layers (i.e., realizing other interactions play a role) and splitting a network into layers (i.e., recognizing distinct kinds of interactions) facilitate cascades. 
Notably, layers unsusceptible to global cascades can cooperatively achieve them if coupled. 
On one hand, this suggests fundamental limitations on predicting cascades without full knowledge of a system's multiplexity; on the other hand, it offers feasible means to control cascades by 
introducing or removing sparse layers in an existing network.
\end{abstract}

\maketitle

When choosing which products to buy, ideas to adopt and movements to join, people are influenced by friends, colleagues, family, and other types of contacts. 
Such influence along multiple channels is frequently non-additive: Just one type of relationship often suffices to convince someone to change behavior~\cite{PadgettAnsell}. 
Banks also interact in many ways---through balance sheet claims, derivatives contracts and reliance on credit lines---which collectively and nonlinearly cause cascades~\cite{ChanLau2010}. 
At a broader scale, countries interact not only through trade~\cite{Lee2011} but also through investment and lending~\cite{Hattori2007}.
Models with just one type of edge \cite{networks} cannot capture this non-additive influence along multiple channels. Instead, one needs graphs that explicitly contain multiple types of edges (or network layers), called \emph{multiplex} networks~\cite{PadgettAnsell,Szell,KMLee}.
Such multiplex networks provide a complementary framework to the growing body of works on interacting and interdependent coupled network systems \cite{Kurant,LeichtDSouza,Buldyrev2010,Gao2012,CDB-sandpile}.

Here we study the impact of such network multiplexity on  cascade dynamics in the threshold model introduced by Watts~\cite{Watts2002}. 
In this stylized model of, for example, contagious behavioral adoption in a social network~\cite{Schelling73,Granovetter}, people join the growing movement if a sufficiently large fraction of their friends have. 
Similarly, banks default if sufficiently many debtor banks default~\cite{GaiKapadia2010}. 
 Specifically, nodes exist in one of two states, active and inactive. 
Each node independently draws a (frozen) threshold $r \in [0,1]$ from a probability distribution $Q(r)$. 
A node of degree $k$ activates if its fraction $m/k$ of active neighbors exceeds its threshold $r$. 
Of particular interest are so-called global cascades, in which a finite fraction of the infinite network becomes activated from a vanishingly small fraction of initially active seeds. 
A key lesson from previous studies is that, for a given distribution of thresholds, network connectivity constrains  global cascades~\cite{Watts2002,GleesonCahalane2007}. 
If it is too sparse, a network lacks a giant component and the connectivity needed for a global cascade; if it is too dense, a network likely cannot surround nodes with sufficiently many active neighbors. Various generalizations have since been introduced and studied~\cite{GleesonCahalane2007,centola,GaiKapadia2010,Hackett2011,Melnik2011}.

\begin{figure}[tb]
\vskip -.2cm
\centering
\includegraphics[width=.88\columnwidth]{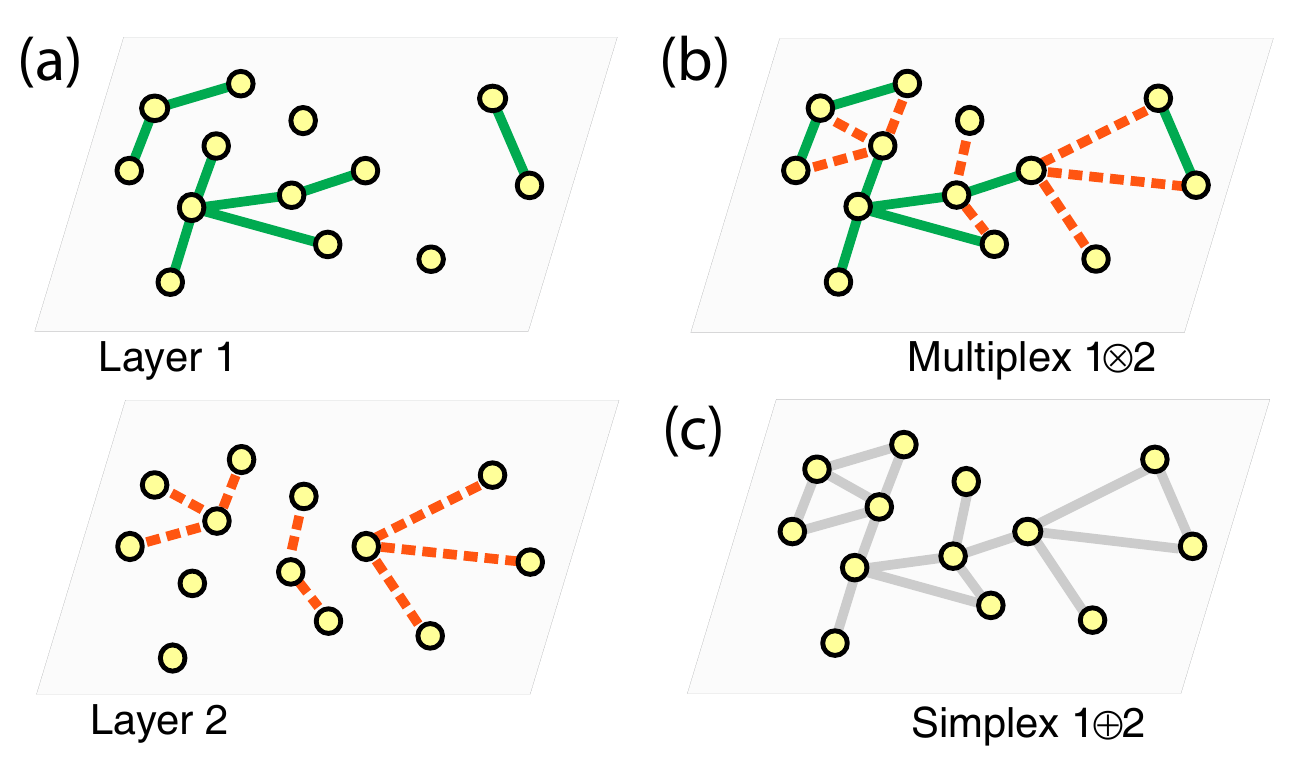} 

\caption{(Color online) Construction of the multiplex 1$\otimes$2 network (b) by combining two layers (a), the solid green (top) and dashed red (bottom) interactions. The simplex 1$\oplus$2 network (c) ignores the types of interaction (solid gray edges).}\label{fig:Fig1-schematic-model}
\end{figure}

In this Rapid Communication we generalize Watts' threshold model \cite{Watts2002} to multiplex networks, in which nodes activate if a sufficiently large fraction of neighbors in {\em any} layer are active. 
To motivate this formulation, note 
that in many situations what matters 
is the influence from one layer alone. For example, a 
large fraction of colleagues recommending a certain 
smartphone application may 
convince someone to use it. Similarly, the default of sufficiently many loans may suffice to depress a country's trade, and vice versa.
To be specific, in a network with two layers (a duplex network), a node with $k_1$ and $k_2$ many neighbors in layers 1 and 2, respectively, with $m_1$ and $m_2$ of those neighbors active, itself activates if $m_1/k_1$ or $m_2/k_2$ exceeds the node's threshold $r$. We denoted this multiplex model the 1$\otimes$2 model. For comparison, we also consider the simplex network that has the same topology but that ignores multiplexity, denoted the 1$\oplus$2 model (Fig.~\ref{fig:Fig1-schematic-model}).

The central result of this Rapid Communication is the greater ease of cascades in multiplex networks. We demonstrate the effect of multiplexity on cascades in two scenarios.  First, given a singe-layer network [denoted Layer 1 in Fig.~\ref{fig:Fig1-schematic-model}(a)], one might realize that another kind of interaction (Layer 2) plays a role. In this case, we combine a second layer (Layer 2) with the existing one to form the duplex network 1$\otimes$2 [Fig.~\ref{fig:Fig1-schematic-model}(b)]. Alternatively, one might realize that a given network in fact consists of multiple channels of non-additive interactions, so the simplex network [Fig.~\ref{fig:Fig1-schematic-model}(c)] is split into two layers to form the 1$\otimes$2 network [Fig.~\ref{fig:Fig1-schematic-model}(b)]. 
We find that both combining layers and splitting into layers facilitate global cascades. 
Layers that in isolation have too much or too little connectivity to achieve global cascades can cooperatively achieve them if they are multiplex-coupled. We demonstrate this analytically and using simulations, and we conclude with generalizations to networks with three or more layers.

We begin by extending the theory of Ref.~\cite{GleesonCahalane2007} to a duplex network with layers of locally tree-like random graphs on the same set of $N$ nodes. Every node has two independent degrees, $k_1$ and $k_2$, equal to its numbers of neighbors in layers 1 and 2, respectively.
The mean fraction of active nodes in the stationary state, called the mean cascade size $\rho$, equals the probability that a randomly chosen node is active. This probability can be obtained by approximating the network as a tree with the chosen node as its root and by considering the cascade of activations toward the root~\cite{GleesonCahalane2007}.  Given the initial seed fraction $\rho_0$, $\rho$ for a duplex 1$\otimes$2 network is given by 
\begin{align}
\rho &= \rho_0 + (1-\rho_0) \sum_{k_1+k_2 \geq 1} p_{k_1}^{(1)} p_{k_2}^{(2)} \times \notag \\
&\qquad \times \sum_{m_1=0}^{k_1} \sum_{m_2=0}^{k_2} B_{m_1}^{k_1}(q_\infty^{(1)}) B_{m_2}^{k_2}(q_\infty^{(2)})  F_{m_1, m_2}^{k_1, k_2}, \label{rhoequation}
\end{align}
where $(q_\infty^{(1)},q_\infty^{(2)})$ is the fixed point of the recursion
\begin{align}
\label{recursion}
\left( \begin{array}{c} q_{n+1}^{(1)} \\ q_{n+1}^{(2)}  \end{array} \right) = \left( \begin{array}{c} g^{(1)}(q_n^{(1)},q_n^{(2)}) \\ g^{(2)}(q_n^{(1)},q_n^{(2)})  \end{array} \right)
\end{align}
begun from $q_0^{(i)} \equiv \rho_0$, where, for distinct $i,j \in \{1,2\}$,
\begin{align}
g^{(i)}(q_n^{(1)},&q_n^{(2)}) = \rho_0 + (1-\rho_0) \sum_{k_i=1}^\infty \sum_{k_j = 0}^\infty \frac{k_i p_{k_i}^{(i)}}{z_i} p_{k_j}^{(j)}\times \notag \\
& \times \sum_{m_i=0}^{k_i-1} \sum_{m_j=0}^{k_j} B_{m_i}^{k_i-1}(q_n^{(i)}) B_{m_j}^{k_j}(q_n^{(j)})  F_{m_1, m_2}^{k_1, k_2}.\label{gequation}
\end{align}
Here $p_{k_i}^{(i)}$ is the degree distribution of layer $i$; $B_n^k(q) \equiv \binom{k}{n} q^n(1-q)^{k-n}$ is shorthand for the binomial distribution; $q_n^{(i)}$ is the probability that a node $n$ steps above the leaves of the tree is activated by its children in the tree, conditioned on its parent in layer $i$ being inactive; $F_{m_1, m_2}^{k_1, k_2}$ is the response function, the chance that a node with $k_i$ neighbors in layer $i$ ($m_i$ of which are active) becomes active; and the factor $k_i p_{k_{i}}^{(i)}/z_i$ in \eqref{gequation} is the probability that a degree-$k_i$ node lies at the end of a randomly chosen edge in layer $i$, where $z_i$ is the mean degree in layer $i$. The response function for the multiplex 1$\otimes$2 model is
\begin{align}
F_{m_1, m_2}^{k_1, k_2} =
	\begin{cases}
		0 & \text{if $\max(m_1/k_1, m_2/k_2) \leq r$,} \\
		1 & \text{if $\max(m_1/k_1, m_2/k_2) > r$.}
	\end{cases}
\end{align}
In this work all nodes have the same threshold $R$ [i.e., $Q(r)=\delta(r-R)$].

\begin{figure}[t]  
\begin{center}
\includegraphics[width=.99\columnwidth]{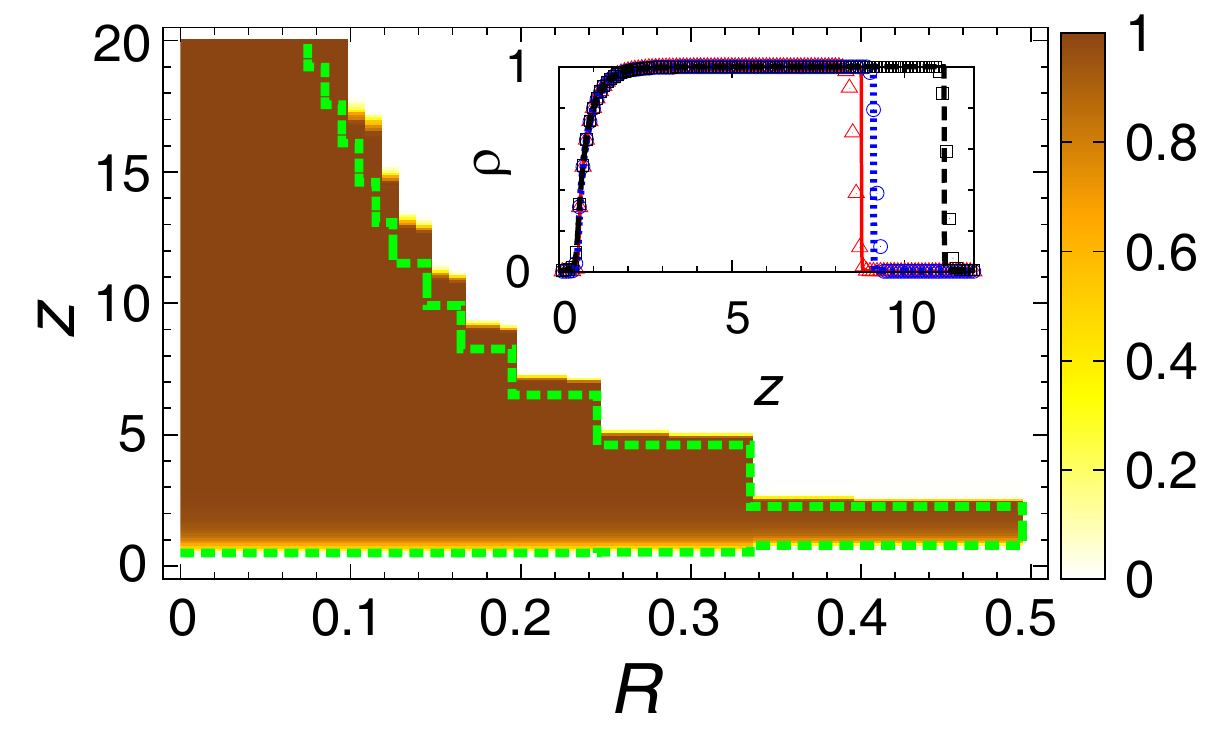}
\vskip -0.3cm
\caption{(Color online) Theoretical cascade boundary given by the first-order cascade condition~\eqref{cascadecondition} (dashed green line) and numerically simulated mean cascade size $\rho$ for $\rho_0=10^{-3}$ (color coded) on a duplex network of $N=10^5$ with ER layers of equal mean degrees $z$. Inset: $\rho$ vs $z$ for threshold $R=0.18$ and different $\rho_0 = 5\times10^{-4}$ (red \textcolor{red}{$\triangle$}), $10^{-3}$ (blue \textcolor{blue}{$\bigcirc$}), $5\times10^{-3}$ ($\square$), obtained from simulations (symbols) and from Eq.~\eqref{rhoequation} (lines).}
\vskip -.5cm
\label{fig:CompareTheorySim}
\end{center}
\end{figure}

To test the validity of the theory, we calculated $\rho$ from \eqref{rhoequation} for duplex \ER (ER) networks~\cite{ER} of layers with the same mean degree $z$. The calculated $\rho$ as a function of $z$ with different seed sizes $\rho_0$ are found to agree well with 
numerical simulations (Fig.~\ref{fig:CompareTheorySim}, inset). As in the single-layer case \cite{Watts2002,GleesonCahalane2007}, for given $R$, global cascades occur for an interval of $z$ between two transitions: a continuous transition at small $z$ for the emergence of global cascades (following the emergence of the giant connected component), and a discontinuous transition for the disappearance of global cascades, the location of which increases with $\rho_0$. The plot of $\rho$ in $(R,z)$-parameter space (Fig.~\ref{fig:CompareTheorySim}, main plot) shows the cascade region, the parameters for which global cascades occur. 

The linear stability of the fixed point $(q^{(1)},q^{(2)})=(0,0)$ of the recursion~\eqref{recursion} as $\rho_0\to0$ gives a sufficient condition for 
global cascades, leading to the so-called first-order cascade condition \cite{Watts2002,GleesonCahalane2007,Hackett2011,Melnik2011}. In the multiplex 1$\otimes$2 case, this condition is that  the maximum eigenvalue of the Jacobian matrix $\mathbf{J}$ of the recursion~\eqref{recursion} at the origin as $\rho_0 \rightarrow 0$ exceeds 1:
\begin{align}
\label{cascadecondition}
\lambda_{\mathrm{max}}(\mathbf{J}) >1.
\end{align}
The 2$\times$2 matrix $\mathbf{J}$ is given by, from (3) and (4),
\begin{align}
\label{Js}
J_{11} = \sum_{k_1 = 1}^{\lfloor{1/R}\rfloor}\frac{k_1(k_1-1) p_{k_1}^{(1)}}{z_1}, \quad
J_{12} = \sum_{k_2 = 0}^{\lfloor{1/R}\rfloor} k_2 p_{k_2}^{(2)},
\end{align}
and similarly for $J_{21}$ and $J_{22}$, where $\lfloor \cdot\rfloor$ denotes the floor function. 
As shown in Fig.~\ref{fig:CompareTheorySim} (main plot), this cascade condition~\eqref{cascadecondition} closely approximates the boundary of the cascade region from simulations, providing a useful approximation to the actual cascade region. Deviations from simulations occur because
~\eqref{cascadecondition} ignores activations by multiple active neighbors. (Including second-order activations improves the agreement
~\cite{GleesonCahalane2007}). For the exact boundary, one must use Eqs.~\eqref{rhoequation}--\eqref{gequation}, as in Fig.~\ref{fig:CompareTheorySim} (inset).

\begin{figure}[b]
\centering
\includegraphics[width=.95\columnwidth]{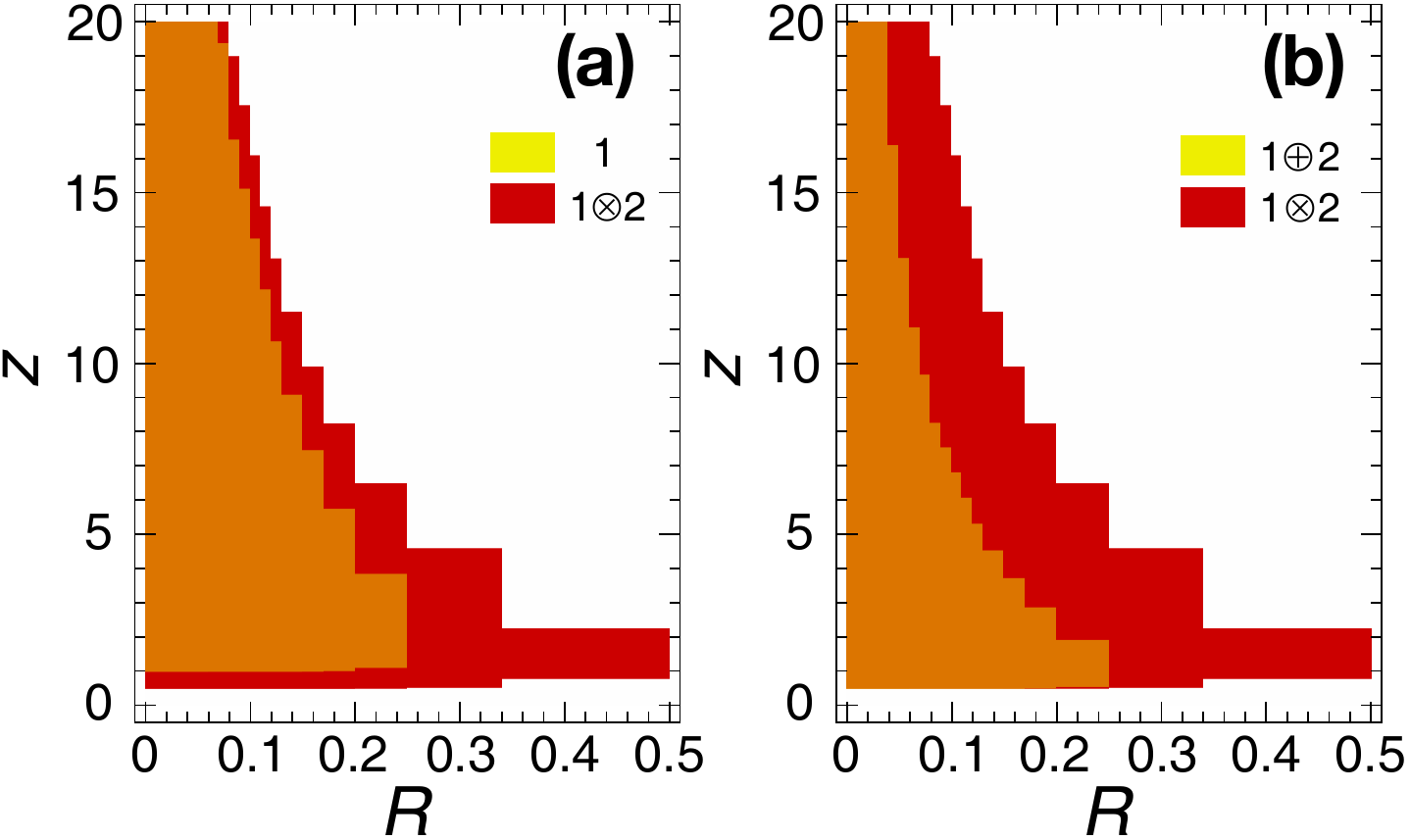} 
\caption{
(Color online) 
Both combining layers (a) and splitting into layers (b) facilitate cascades. (a) Cascade regions from cascade condition~\eqref{cascadecondition} for single-layer (yellow, brighter) and duplex (red, darker) network with ER layers of equal mean degree $z$. (b) Same plots for a simplex ER network of mean degree $2z$ (yellow, brighter) and corresponding duplex network (red, darker). Overlapping regions appear orange (medium brightness).} 
\vskip -.2cm
\label{fig3}
\end{figure}

The following interpretation of the cascade condition elucidates how multiplexity facilitates cascades. 
The matrix $\mathbf{J}$ can be identified with the mean reproduction matrix of a two-type branching process~\cite{athreya}, with $J_{ij}$ representing the mean number of offspring of type-$j$ branching from a node born through a type-$i$ branching. Here
a type-$i$ branching corresponds to activation along the layer $i$ by a single active node. The cascade condition is the supercriticality condition for this two-type branching process. This branching process approximates the progression of actual cascades by ignoring activations by multiple active neighbors.
Cascades in a single-layer network are approximated by a single-type branching process, so the cascade condition becomes $J_{ii}(R,z)>1$~\cite{Watts2002}. 
For the 1$\otimes$2 network, the cascade condition reads $\lambda_{\mathrm{max}}=\frac{1}{2}\left[(J_{11}+J_{22})+\sqrt{(J_{11}-J_{22})^2+4J_{12}J_{21}}\right]>1$. 
Since this $\lambda_{\mathrm{max}}\ge\max(J_{11},J_{22})$, the 1$\otimes$2 cascade region contains the cascade regions of either of its layers in isolation, for any degree distributions. 
Thus what enlarges the cascade regions is the presence of an additional activation channel, represented by the off-diagonals $J_{ij}$. 
The two types of activation channels promote each other's activations in a cooperative, positive-feedback manner, collectively facilitating cascades in multiplex networks.

As an explicit example, we consider duplex networks with ER layers of equal mean degrees under the two scenarios of multiplexity discussed above. First, a second ER layer with the same mean degree $z$ is combined with an existing ER layer with mean degree $z$ to form the 1$\otimes$2 network (of total mean degree $2z$), akin to considering lending as well as trading relationships among countries in the global economic system. 
Second, an ER graph with mean degree $2z$ is randomly split into two ER layers with equal mean degrees $z$ to form the 1$\otimes$2 network, akin to distinguishing social influence among colleagues and among friends in a social network. Explicit evaluation of the cascade condition (5) yields, in both cases, an enlarged cascade region for the 1$\otimes$2 network (Fig.~\ref{fig3}). 

\begin{figure}
\includegraphics[width= 0.45\textwidth]{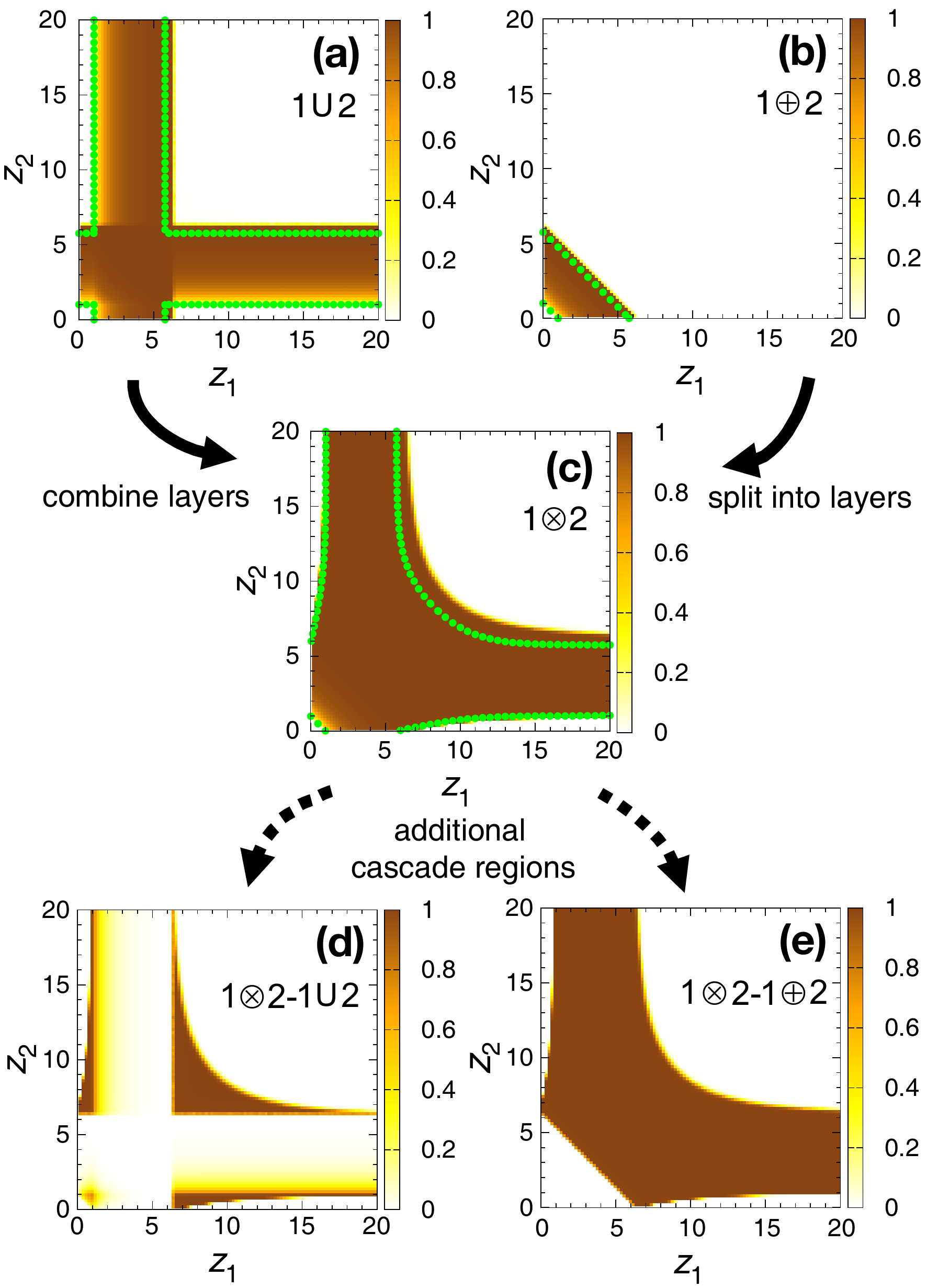}
\caption{
(Color online) Cascade regions with $R=0.18$ for (a) the union 
 of two independent ER layers with mean degrees $z_1$ and $z_2$ (denoted 1$\cup$2), (b) the simplex (1$\oplus$2) ER network with mean degree $z_1+z_2$, and (c) the duplex (1$\otimes$2) network with ER layers of mean degrees $z_1$ and $z_2$. The dotted green lines are the cascade boundary obtained from Eq.~\eqref{cascadecondition}; numerically-simulated $\rho$ ($\rho_0=10^{-3}$) is color-coded. 
Also shown are additional cascade regions in the multiplex cases for (d) combining and (e) splitting.
 }
\label{fig:different-z}
\end{figure}

A multiplex network with statistically distinct layers supports even more nontrivial cascades. 
Both combining two ER layers with mean degrees $z_1$ and $z_2$ [Figs.~\ref{fig:different-z}(a),~\ref{fig:different-z}(c), and~\ref{fig:different-z}(d)] and splitting an ER network into two layers with mean degrees $z_1$ and $z_2$ [Figs.~\ref{fig:different-z}(b),~\ref{fig:different-z}(c), and~\ref{fig:different-z}(e)] enlarges the cascade region. 
The additional cascade regions in the multiplex case [Figs.~\ref{fig:different-z}(d) and~\ref{fig:different-z}(e)] highlight the cooperative effect of multiplexity. For these parameters, each layer is too sparse or too dense to  achieve global cascades, but they   cooperatively achieve them when multiplex-coupled. Of particular interest is when one of $z_1, z_2$ 
is too small ($<1$), the other too large ($\gtrapprox 6$) to support global cascades in isolation. 
This presents a way to control cascades in a system by 
introducing or removing sparse layers below the percolation threshold, which may be more feasible to implement than perturbing the existing, dense network.

We verified a similar enlargement of cascade regions for networks with layers of broad degree distributions by using the static model of scale-free graphs, which allows a variable mean degree~\cite{Goh2001}. Furthermore, we checked that short loops introduced by multiplexity can be neglected~\cite{Hackett2011} for the large, sparse networks considered here. 

\begin{figure}[t]
\centering
\includegraphics[width=.9\columnwidth]{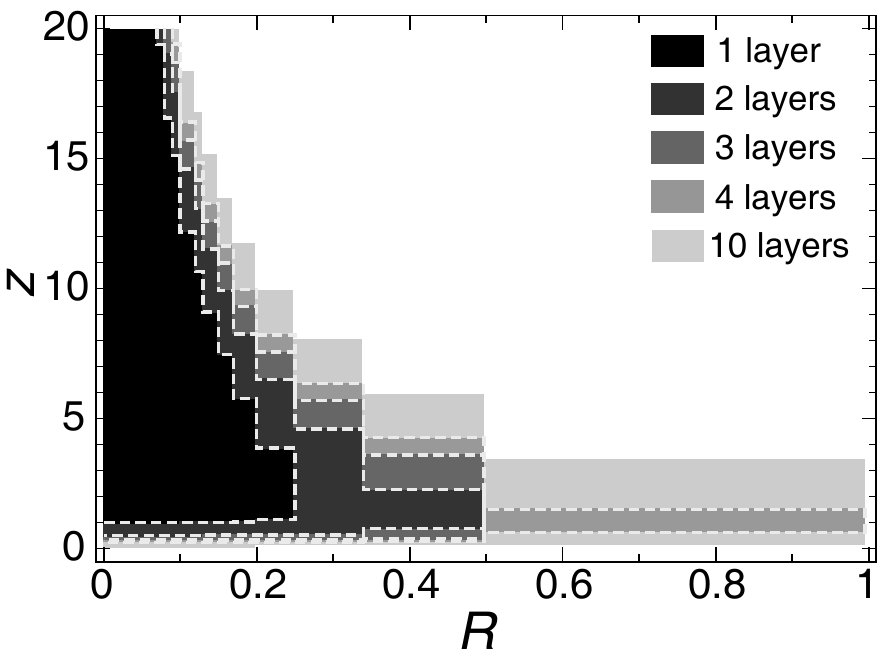}
\caption{
Cascade condition of $\ell$-plex networks of ER layers each with mean degree $z$, for $\ell=1, 2, 3, 4$ and $10$. Note that the cascade region extends to $R\ge1/2$ for $\ell\ge 4$. We show white, dashed lines on boundaries of regions for visual clarity.}
\label{fig:fourlayers}
\vskip -.2cm
\end{figure}

Finally we generalize to multiplex networks with $\ell >2$ layers. Extending Eqs.~\eqref{rhoequation}--\eqref{Js} to $\ell$ layers is straightforward. Combining more layers 
further facilitates cascades for larger $z$ and $R$ (Fig.~\ref{fig:fourlayers}). Notably, introducing a fourth layer permits global cascades even for thresholds $R \geq 1/2$, which Morris~\cite{morris} proved cannot occur in simplex networks. This can be understood from the cascade condition as follows. For $R\ge1/2$, only activations of degree-1 nodes contribute to the cascade condition (5) and (6). For multiplex networks with $\ell$ ER layers with equal mean degree $z$, the first-order cascade condition for $R\ge1/2$ thus becomes $(\ell-1) z e^{-z}>1$, which can be satisfied by a non-empty interval of $z$ for $\ell\ge4$. This suggests that even people 
difficult to persuade ($R\geq 1/2)$ to buy a new device, for example, may all buy one if they participate a little ($z\approx1$) 
in many social spheres ($\ell \geq 4$).

To conclude, the interplay among multiple kinds of interactions---the multiplexity---can generically increase a network's vulnerability to global cascades in a threshold model. Interestingly, layers unsusceptible to cascades can cooperatively become susceptible when coupled. The impact of multiplexity on network dynamics is expected to be widespread \cite{LeichtDSouza,Buldyrev2010,Gao2012,CDB-sandpile,Lee-sandpile}. In other binary-state, monotonic, threshold cascade models such as bootstrap percolation~\cite{Baxter2010}, a similar analysis can be readily applied, while its impact on more complicated dynamics~\cite{dynamics} remains largely unexplored. Our results suggest a double-faceted picture for cascade prediction and control. 
On one hand, one faces fundamental limitations on predicting cascades without full knowledge of a system's multiplexity; on the other hand, multiplexity offers a feasible tactic to enable or hinder cascades by introducing or removing sparse layers, respectively. 
For instance, advertising may become more effective with every new medium, while banks may grow more vulnerable with every new lending mechanism.

\begin{acknowledgments}
This work was supported in part by the Mid-career Researcher Program (No.\ 2009-0080801) and Basic Science Research Program (No.\ 2011-0014191) through NRF grants funded by the MEST of Korea; K.-M.L. was funded by Global Ph.D. Fellowship Program (No.\ 2011-0007174) through NRF, MEST; 
and C.D.B. was funded by NSF EAPSI, Award 1107689, and DTRA Basic Research Award HDTRA1-10-1-0088.
\end{acknowledgments}



\begin{thebibliography}{99}

\bibitem{PadgettAnsell} J.\ F.\ Padgett and C.\ K.\ Ansell, Am. J. Sociol. {\bf 98}, 1259 (1993).
\bibitem{ChanLau2010} J. A. Chan-Lau, IMF Working Paper No. 10/107, Washington, DC, April 2010 [\url{http://ideas.repec.org/p/imf/imfwpa/10-107.html}].
\bibitem{Lee2011} K.-M. Lee {\it et al.}, PLoS ONE {\bf 6}, e18443 (2011).
\bibitem{Hattori2007} M. Hattori and Y. Suda, Research on global financial stability: the use of BIS international financial statistics {\bf 29}, 16 (2007). 
\bibitem{networks} S. Boccaletti {\it et al.}, Phys. Rep. {\bf 424}, 175 (2006).
\bibitem{Szell} M. Szell, R. Lambiotte, and S. Thurner, Proc. Natl. Acad. Sci. U.S.A. {\bf 107}, 13636 (2010).
\bibitem{KMLee} K.-M. Lee {\it et al.}, New J. Phys. {\bf 14}, 033027 (2012).
\bibitem{Kurant} M. Kurant and P. Thiran, Phys. Rev. Lett. {\bf 96}, 138701 (2006).
\bibitem{LeichtDSouza} E. A. Leicht and R. M. D'Souza, arXiv:0907.0894 (2009).
\bibitem{Buldyrev2010} S. V. Buldyrev {\it et al.}, Nature (London) {\bf 464}, 1025 (2010).
\bibitem{Gao2012} J. Gao, S. V. Buldyrev, H. E. Stanley, and S. Havlin, Nat. Phys. {\bf 8}, 40 (2012) and references therein.
\bibitem{CDB-sandpile} C. D. Brummitt, R. M. D'Souza, and E. A. Leicht, Proc. Natl. Acad. Sci. U.S.A. {\bf 109}, E680 (2012). 
\bibitem{Watts2002} D.\ J.\ Watts, Proc.\ Natl.\ Acad.\ Sci.\ U.S.A. {\bf 99}, 5766 (2002).
\bibitem{Schelling73} T. C. Schelling, J. Conflict Resolution {\bf 17}, 381 (1973).
\bibitem{Granovetter} M. Granovetter, Am. J. Sociol. {\bf 83}, 1420 (1978).
\bibitem{GaiKapadia2010} P. Gai and S. Kapadia, Proc. R. Soc. A {\bf 466}, 2401 (2010).
\bibitem{GleesonCahalane2007} J. P. Gleeson and D. Cahalane, Phys. Rev. E {\bf 75}, 56103 (2007).
\bibitem{centola} D. Centola, V. Egu\'iluz, M. Macy. Physica A {\bf 374}, 449 (2007).
\bibitem{Hackett2011} A. Hackett, S. Melnik and J. P. Gleeson, Phys. Rev. E {\bf 83}, 056107 (2011).
\bibitem{Melnik2011} S. Melnik, J. A. Ward, J. P. Gleeson, and M. A. Porter, e-print arXiv:1111.1596.
\bibitem{ER} P. Erd\H{o}s and A. R\'{e}nyi, Publ. Math. Inst. Hung. Acad. Sci. {\bf 5}, 17 (1960).
\bibitem{athreya} K. B. Athreya and P. E. Ney, {\it Branching processes} (Springer-Verlag, Berlin, 1972). 
\bibitem{Goh2001} K.-I. Goh, B. Kahng, and D. Kim, Phys. Rev. Lett. {\bf 87}, 278701 (2001).
\bibitem{morris} S. Morris, Rev. Econ. Stud. {\bf 67}, 57 (2000).
\bibitem{Lee-sandpile} K.-M. Lee, K.-I. Goh, and I.-M. Kim, J. Korean Phys. Soc. {\bf 60}, 641 (2012). 
\bibitem{Baxter2010} G. Baxter, S. Dorogovtsev, A. Goltsev, and J. Mendes, Phys. Rev. E {\bf 82}, 011103 (2010).
\bibitem{dynamics} A. Barrat, M. Barth\'elemy, and A. Vespignani, {\it Dynamical processes on complex networks} (Cambridge University Press, Cambridge, 2008).

\end{thebibliography}
\end{document}